# Elemental Stoichiometry as an Ecological Biosignature with Applications to Life Detection

Pilar C. Vergeli[1†], Cole Mathis[2], John F. Malloy[1], L. Felipe Benites[4], Christopher P. Kempes[3], Elizabeth Trembath-Reichert[1], Hilairy E. Hartnett[1,5‡], Sara I. Walker[1,3,4*]
[1]School of Earth and Space Exploration, Arizona State University, Tempe, AZ
[2]School of Complex and Adaptive Systems, Arizona State University, Tempe, AZ
[3]The Santa Fe Institute, Santa Fe, NM
[4]The Beyond Center for Concepts in Fundamental Science, Arizona State University, Tempe, AZ
[5]School of Molecular Sciences, Arizona State University, Tempe, AZ

* Author for correspondence: sara.i.walker@asu.edu

† Present address: Carnegie Institution for Science, Caltech, Pasadena, CA 91125, USA.
‡ Present address: School of Oceanography, University of Washington, Seattle, WA 98195, USA.

## Abstract

The vast chemical space of possible small molecules, estimated at $10^{60}$ compounds for molecules composed of just C, N, O, and S, is only sparsely occupied by biology. We propose that where life selects molecules within this space constitutes a detectable ecological signature: a fingerprint not of specific compounds, but of the statistical structure of elemental composition across molecules sampled from ecological systems. Here we introduce a framework combining Van Krevelen diagrams and element scaling laws to characterize the elemental composition of regions of chemical space occupied by biological systems and contrast them with other chemical systems. Applying this framework to 11,834 microbial metagenomic samples, we show that microbial metabolisms occupy a region of chemical space, which is enriched in heteroatoms such as P, S, N, and O relative to C, shifted toward higher O:C and H:C ratios. We observe sublinear element scaling with system size, yielding insights into how elemental constraints dictate how biological systems occupy chemical space. These patterns are distinct from a sample of ~18,000 compounds from the comprehensive Reaxys synthetic chemical database. Critically, datasets from molecules detected in planetary science mission data occupy statistically distinct regions from both terrestrial biological and Reaxys distributions, demonstrating that with standardized methods for data collection, the approach could be developed to discriminate biotic from abiotic chemical signatures in small molecule data from planetary science missions. Our work shows how a combination of Van Krevelen fingerprinting and elemental scaling laws can provide a new class of ecological biosignatures for life detection leveraging mass spectrometric data from planetary missions, which could generalize beyond Earth's specific biochemistry.

# Main

One of the central challenges in life detection is determining how to recognize alien life if its chemistry is different from that of Earth's biology. Addressing this challenge requires identifying features of biochemistry that may generalize beyond Earth-specific biology. This is challenging given how chemical space, the property space spanned by all possible compounds consistent with a given set of physical construction rules and chemical conditions, is astronomically large (1–3). Considering only compounds composed of the small set of elements including carbon, hydrogen, nitrogen, oxygen, and sulfur yields a rough combinatorial estimate of $10^{60}$ compounds (3, 4). By contrast, across all known metabolisms there are only on the order of ~$10^{4}$ verified biochemical compounds (5). Biology, in other words, occupies a vanishingly small corner of possible chemical space. Here we aim to characterize the properties of biochemistry within chemical space, to develop data-driven first principles approaches to life detection.

Recent work has considered whether molecules carry structural signatures of the evolutionary and selective processes necessary to generate them. The molecular assembly index, a measure of the minimum number of steps required to build a molecule from basic constituents, gives a quantitative structure to chemical space, which allows measuring molecular complexity in the lab with mass spectrometry, NMR, and infrared spectroscopy and relates these measurements to the degree of selection required for the molecule's production (6, 7). This theory suggests molecules with high molecular assembly indices are too complex to arise by chance in abiotic systems, allowing defining an assembly threshold where molecules become definitive biosignatures (8): existence of abundant high assembly index implies a history of iterative selection and construction within chemical space that is, to date, only observed in living or life-derived processes. Metrological tests of the theory across diverse sample types using mass spectrometry place the threshold at abundant molecules with assembly index ~15 (6). However, current flight-ready instrumentation, including the mass spectrometers aboard planetary missions, typically returns molecular formula data rather than full structural information, and these data are limited to small molecular weight molecules (9, 10). These limitations place nearly all potentially detectable molecules on planetary science missions below the assembly threshold. While new methods are being developed to assess molecular assembly index information from mission-relevant data (11, 12), small molecule biochemistry may encode other indications of evolutionary history related to assembly theoretic principles of selection within chemical space. Other approaches to molecular biosignatures based on gas chromatography-mass spectrometry (GC-MS) data have leveraged advances in machine-learning (13, 14) and still others have applied statistical measures to distinguish molecular diversity in biological and abiotic systems (15). However, these approaches risk over-fitting to Earth-life and do not provide the mechanistic insights that would allow generalizing our understanding of selection and evolution beyond known biochemistry. Thus, here we adopt a complementary approach to the emerging framework of molecular assembly, aimed at extracting evidence for living processes from evidence of selection within the chemical space of small molecules.

The positioning of biology within chemical space is not arbitrary but reflects constraints on how living systems acquire and organize elements into functional molecules, including catalysts, energy carriers, structural materials, and information-bearing polymers. Life's molecular machinery is dominated by the six bioessential elements, carbon, hydrogen, nitrogen, oxygen, phosphorus, and sulfur (CHNOPS), which occur in tightly constrained stoichiometric ratios shaped by both the chemical properties of the elements themselves and evolutionary history (16–19).

However, macromolecular elemental composition, being strongly influenced by shared ancestry, may not generalize across alternative evolutionary trajectories, motivating a focus on small molecule metabolism as potentially a more universal and tractable target.

While an abundant complex molecular species (like DNA or RNA) may provide surefire signs of its evolutionary origin, small molecules are known to form in abundance abiotically. Thus, rather than asking which specific molecules are present in a sample, assessing biogenicity for small molecule chemistry will require an approach looking at patterns of selection embedded in the composition of an ensemble of distinct molecular species in a sample. We ask whether the statistical pattern of elemental composition across a sample of molecules carries a recognizable imprint of its ecological origins. Our framing is grounded in the practical realities of data. Experiments on extraterrestrial samples will, in the near term, be unlikely to differentiate distinct lineages, and instead will likely represent ecological communities (if life is present), much like environmental microbial sampling on Earth. If biological systems consistently occupy a characteristic region of chemical space, the geometry of that region could provide a life detection signal that is independent of the exact compounds involved and, potentially, independent of the specific biochemistry and selective history underlying them.

Small molecules, which serve as substrates, products, and cofactors of enzymatic reactions, are chemically diverse, environmentally responsive, detectable with flight-ready instrumentation, and may reflect metabolic organization independent of specific evolutionary histories (20–23, 9, 10). Within this context, chemical space, defined by molecular properties independent of synthesis or abundance, provides a quantitative framework for identifying statistical regularities in element use.

By examining where biological molecules fall relative to other chemical systems, we can detect patterns of elemental stoichiometry shaped by the bioelement distributions that define biogeochemical niches (elementomes) representing encoded evolutionary selection in a measurable and scalable form (24, 25). We develop this framework for molecular chemical space.

We investigate these element stoichiometry patterns using two complementary analytical approaches. The first is Van Krevelen diagrams, originally developed for coal chemistry and now widely used in metabolomics, which map, or fingerprint, elemental ratios of individual compounds into a two-dimensional coordinate space (H:C versus P:C, S:C, N:C, or O:C), revealing characteristic clustering by compound class (26, 27). The second is a method newly introduced to this work, where we study element scaling laws relating the total abundance of each bioessential element to carbon across entire biological systems, capturing emergent stoichiometric constraints that only become apparent at the system level rather than the molecular level (28).

We apply this framework to three classes of chemical data: a large-scale metagenomics dataset spanning over 11,000 microbial communities drawn from diverse environments (Environmental Microbial Space), the full set of known biochemical compounds cataloged in the CBR-db database (Biochemical Space), and a matched sample of ~18,000 compounds from the Reaxys chemical database (Synthetic Chemical Space) representing synthesized chemical diversity (29–31). We show that microbial metagenomes occupy a statistically distinct and environmentally universal region of chemical space, which is enriched in heteroatoms relative to carbon and exhibits sublinear element scaling. We further demonstrate that the same framework can be applied to mission data, such as returned Bennu asteroid samples (32, 33). Although the datasets differ in curation methods and collection standards, the framework we present illustrates the method's applicability to both terrestrial and extraterrestrial chemistry, motivating future work in sample standardization for purposes of life detection. Taken together, our results suggest that Van Krevelen fingerprinting and element scaling laws could enable defining ecological molecular biosignatures: detectable imprints of biological selection in the elemental stoichiometry of small molecule chemistry that may generalize to ecologies beyond Earth's specific biochemistry.

# Methods

**Acquiring metagenomic data**

We retrieved metagenomic data from the Department of Energy Joint Genome Institute's Integrated Microbial Genomes and Microbiomes (DOE-JGI IMG/M) database in September of 2023 (34, 35, See **SI Section 1**). We examine the data within the subset of genes for which there is functional annotation (See **SI Section 1**). We acquired 18,824 metagenomes processed through the DOE-JGI IMG/M annotation pipeline and filtered the database with a text-mining Python script (36). We filtered metagenomic data to remove outliers in size and samples that did not include sufficient functional annotation (See **SI Section 2**). For each metagenome, we collected the set of DOE-JGI IMG/M EC annotations for enzyme catalyzed reactions for each metagenome. Six enzyme classes were used for our analysis: oxidoreductases (EC 1), transferases (EC 2), lyases (EC 3), hydrolases (EC 4), isomerases (EC5), and ligases (EC 6). We did not include enzymes belonging to the group translocases (EC 7) due to the inclusion of reactions that are linked to other EC classes (37). In the filtering pipeline, we removed enzymes with partial names, (example: 1.1.1.-), enzymes associated with glycans, metagenomes at the extremes of percent functional coding genes (<10% or >90%), and gene sizes that fell <13,540 genes leaving the smallest metagenomes to potentially contain at ~10 individual archaea or bacteria, based on the genome size of *Pelagibacter ubique* with the smallest number of genes (n=1354) for a free-living microorganism (38). This yielded a dataset of 14,602 metagenomes (See **SI Section 2**).

For each metagenome, we obtained metadata from the Genome OnLine Database v. 10 (GOLD) that adheres to metadata standards as defined by the Genomics Standards Consortium (39, see **SI Section 1**). Our metadata included ecotype descriptors (e.g., latitude/longitude, habitat, ecosystem subtype, ecosystem type, and ecosystem category), genome statistics (e.g., total number of base pairs, the total number of genes, and the number of protein coding genes, etc.). We excluded engineered and host-associated entries and focused our analysis on environmental metagenomes so that our dataset contained metagenomes under more direct environmental selection pressure. We filtered our dataset to include "Aquatic" and "Terrestrial" ecosystem categories, excluding the ecosystem category 'Air'. Finally, we only examined metagenomes for which there were more than 100 entries per ecosystem type for robust statistical representation in further analysis. This yielded our final, curated dataset of 11,834 metagenomes for analyses; this is 63% of the original raw metagenomic data obtained from DOE-JGI. We refer to this curated and filtered metagenome compound data as "Environmental Microbial Space" in what follows, to designate it as a representative sample of the chemical space occupied by environmentally influenced microbial metabolisms.

**Identifying patterns in systems of compounds**

We use two different statistical approaches to examine element patterns: Van Krevelen diagrams to explore element distribution and organization within systems, and element scaling laws to explore elementomes across systems. These two approaches complement each other by providing insights into patterns within specific systems and patterns across systems. We apply these analyses to two sets of biological data, designated "Biochemical Space" and "Environmental Microbial Space". Here "Biochemical Space" refers to the subset of chemical space occupied by small molecule biochemistry inclusive of all known metabolisms: we model this using all compounds (n = 18,716) cataloged in CBR-db. CBR-db is a biochemical reaction database curated from the Kyoto Encyclopedia of Genes and Genomes (KEGG) and ATLAS of Biochemistry (ATLAS) databases, the latter of which includes hypothesized biochemical compounds not yet verified within organisms (31). By contrast "Environmental Microbial Space" considers those subsets of Biochemical Space confirmed to be occupied by microbial community metabolisms from terrestrial and aquatic ecosystems, which we model with the filtered metagenomic compound data. Environmental Microbial Space compounds are compiled through the link between EC number and compounds used in the reaction the EC number represents. Each EC number is matched with a reaction number and the compounds comprising that reaction in the CBR-db database. We use the list of EC numbers per metagenome to generate the list of compounds used in each inferred enzymatic reaction encoded. Using this list of compounds per metagenome, we obtain compound formulas and thus, element counts and ratios per formula.

Van Krevelen diagrams we're traditionally used to display compound stoichiometry with H:C on the y-axis and O:C on the x-axis, defining a two-dimensional chemical space to distinguish the ages of kerogen samples for oil production (26, 40). This approach is broadly applicable to any set of compounds, whether recovered from a metagenome or an environmental sample, and it can be implemented with data derived from samples taken in other planetary environments. Recent work using Van Krevelen analysis has extended to the examination of compound diversity and reaction processes in environmental samples such as aerosols and soils (41, 42). In Van Krevelen coordinate space, compound stoichiometries reveal patterns defined by groups of compounds with similar structural characteristics (**Figure 1B,** See **SI Section 4**). We use Gaussian kernel density estimation (KDE) to characterize the joint distribution of compounds in Van Krevelen space and we computed $10^{th}$-90th percentile ranges for element:C and H:C ratios to summarize the spread of compounds along each axis using the SciPy software package (43). This comparative analysis enables statistical examination of elemental stoichiometry in Biochemical Space and Environmental Microbial Space.

We use scaling analysis to observe stoichiometric features across biological systems. In our analyses, we are interested in scaling in elemental stoichiometry as a window in the structure of the chemical space selected by life, so we use the number of carbon atoms totaled across all unique compounds in a sample as our independent variable (e.g., chemical space system size), and the number of atoms of another elemental type as the dependent variable. This allows us to determine scaling within chemical space, e.g., how are carbon and other elements distributed across molecules as the number of total atoms in the chemical space increases. Scaling laws are quantified by a power-law fit $(y = ax^b)$, with x as the number of carbon atoms in a sample and y as the total count of another element in the same sample, where the exponent $b$ describes the scaling behavior. Scaling fits ($b$) were determined through maximum likelihood estimation using a 95% confidence interval (CI). A 95% CI that is equal to or crosses the value of 1 is considered linear. A sublinear slope is defined by a slope <1 and an upper CI <1. A superlinear slope is defined by a slope >1 and a lower CI >1. We use the scaling behaviors to identify system-level dependencies that may be similar or different across biological and chemical samples.

We summarize how the approach of combining Van Krevelen diagram and scaling analyses can uncover stoichiometric patterns within and across systems, revealing potentially universal system-scale constraints, see **Figure 1**. Both approaches rely only on molecular formulas; therefore, this framework can be applied to any system where molecular formula data can be acquired such as mass spectrometric data from environmental samples or extraterrestrial chemical datasets.

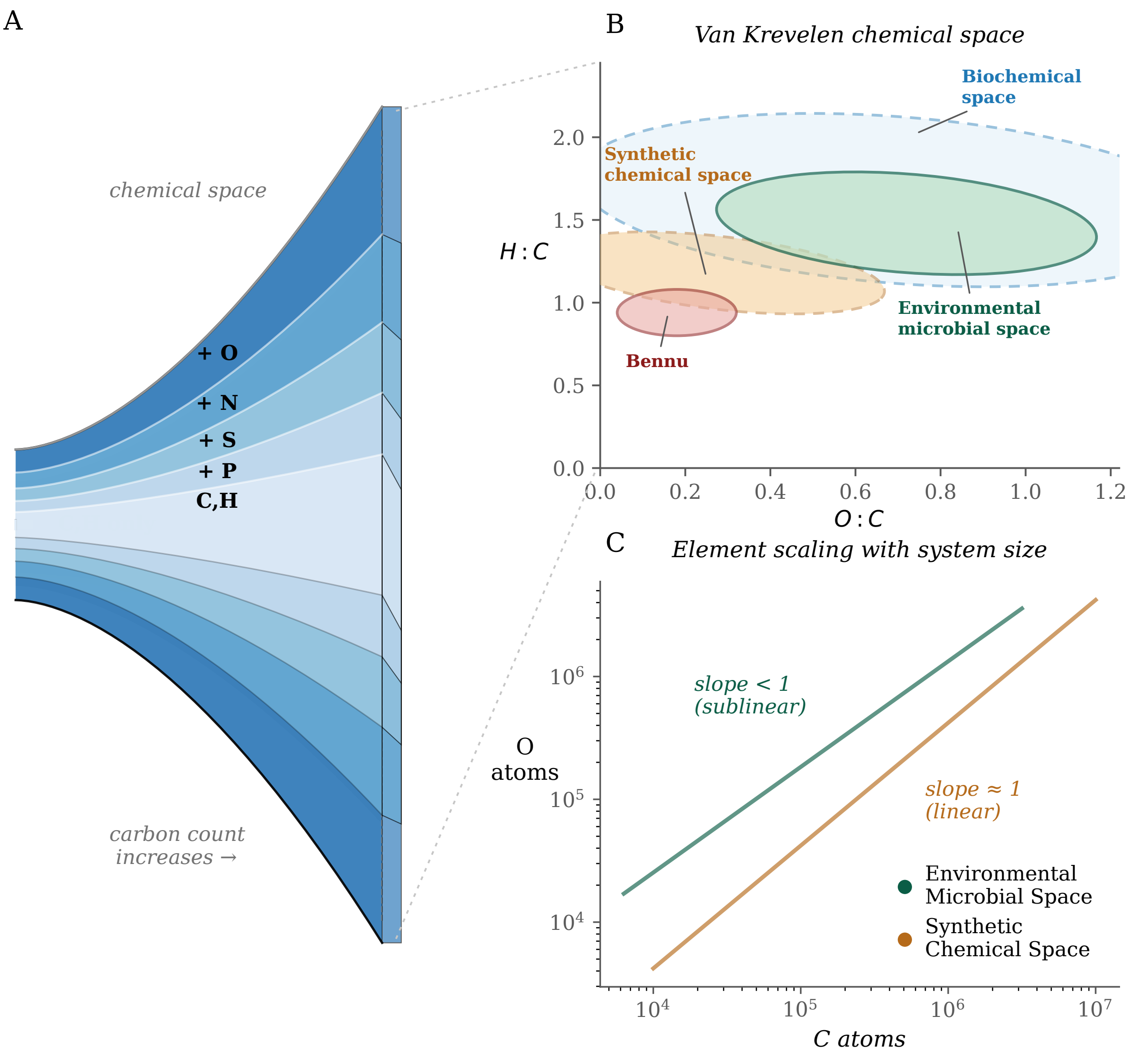

**Figure 1: Element stoichiometry of chemical space across systems and scales.** (A) Conceptual schematic illustrating the expansion of chemical space with increasing elemental diversity: as the number of carbons per molecule increases, the number of possible molecular configurations increases at least exponentially. Carbon, and hydrogen serve as the basic requirements for molecular structures, and the concentric cones visualize how as elemental stoichiometry is diversified with incorporation of heteroatoms (P, S, N, O) the size of chemical space expands. (B) Occupied chemical space regions can differentiate systems with different constraints and selective histories (Synthetic Chemical, Biochemical, Environmental Microbial, and Bennu) by projecting data into Van Krevelen coordinates (O:C vs H:C). (C) Scaling of element stoichiometry across systems, shown here with oxygen atoms versus carbon atoms counted cumulatively across all molecules constituting the chemical space of a sample. Environmental Microbial Space exhibits sublinear scaling (slope < 1), indicating constrained incorporation of oxygen relative to carbon as the size of the chemical space (counted in C atoms) within a sample increase, whereas Synthetic Chemical Space follows linear scaling. Together, these concepts link within-system stoichiometric structure (panel B) to multi-system stoichiometry structure observed through scaling behavior (panel C).

**Sampling Reaxys to explore Synthetic Chemical Space**

To provide a comparison for Environmental Microbial and Biochemical Spaces, we also explore data related to small molecule chemistry that is synthesizable by biological systems or in the chemistry lab. For a synthetic chemical dataset, we generated randomized compound samples from the Reaxys database (see **SI Section 4**). Reaxys is one of the world's largest chemical databases, including data on over 250 million compounds (3, 44). While most of chemical space remains uncharacterized due to its astronomical size (3, 45), Reaxys data allows constructing diverse groups of compound stoichiometry by sampling one of the largest chemical inventories of known chemical structures, comprised of compounds from not just biochemical pathways, but also pharmaceutical, industrial, toxicological, and synthetic pathways (29). We sampled Reaxys compounds within the molecular-weight range constrained to match that of compounds in CBRdb (<2682 Da). We further limited this set of compounds to include C, N, O, P, S, generating a set of 18,000 Reaxys compounds that we refer to as Synthetic Chemical Space hereafter. Only 2 compounds overlapped between Synthetic Chemical Space and Biochemical Space (See **SI Figure 3**), thus in terms of chemical species represented in the data, these datasets are effectively unique. We generated 10,000 compound sets by randomly selecting compounds from Synthetic Chemical Space compounds such that each set size (number of compounds) was comparable to the metagenomic compound data set sizes (e.g. the number of metabolites encoded across the diverse metagenomes).

# Results

### Element ratios of compounds

We examined the overall distributions of elemental ratios (P:C, S:C, N:C, and O:C) across the three chemical spaces using datasets of unique compounds from Reaxys (**Synthetic Chemical Space**; n = 18,000), compounds cataloged in CBR-db (**Biochemical Space**; n = 18,716), and compounds identified across 11,834 metagenomes (**Environmental Microbial Space**; n = 3,743), see **Figure 2** and **Table 1**. The minimum ratio values for P:C, S:C, N:C, and O:C are consistent across the spaces. For P:C ratios, Biochemical Space and Environmental Microbial Space are comparable, both exhibiting higher medians and 90$^{th}$ percentile values than Synthetic Chemical Space. In contrast, S:C ratios exhibited decreasing maximum values from Synthetic Chemical Space to Environmental Microbial Space, while median values remained similar across spaces, and both Biochemical and Environmental Microbial molecules retained higher 90$^{th}$ percentile values than Synthetic Chemical Space. For N:C and O:C ratios, median values increase progressively moving from Synthetic Chemical Space, with the smallest values, to Environmental Microbial Space, with the largest values. Biochemical space exhibited the highest maximum O:C ratio (8.0), while the Environmental Microbial Space O:C ratio exhibited the highest 10$^{th}$ percentile, median, and 90$^{th}$ percentile values.

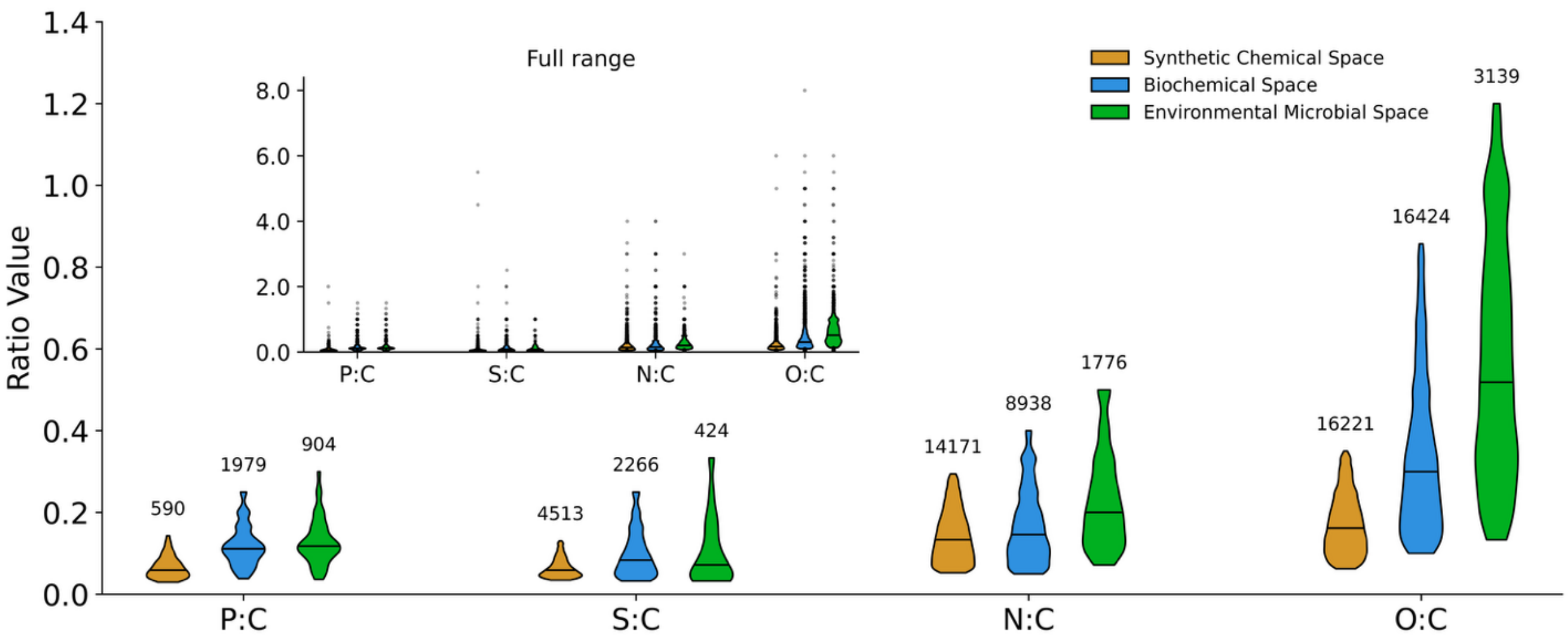

**Figure 2: Element ratios of compounds distinguish different subsets of chemical space.** Carbon-normalized violin plots of elemental ratios for molecules in Synthetic Chemical Space (orange), Biochemical Space (blue), and Environmental Microbial Space (green). The main panel displays a zoomed view spanning the $10^{th}$-$90^{th}$ percentile range of the data, with medians shown as black horizontal lines. Violin widths represent the density distribution of elemental ratios for molecules in each group. Numerical values above each violin indicate sample sizes. (Inset) Full-range violin plots showing the complete distribution of elemental ratios. See **SI Section 4** for distribution summary statistics.

We next compared the compound stoichiometry patterns of C, H, P, S, N, and O in Van Krevelen coordinate space representing Synthetic Chemical Space, Biochemical Space, and Environmental Microbial Space, respectively. We projected our data into Van Krevelen diagrams, mapping P:C, S:C, N:C, and O:C on the x-axis vs. H:C on the y-axis, see **Figure 3**. The coordinate space in a Van Krevelen diagram analysis enables the identification of compound type clustering normalized to carbon, as well as potential reaction processes within chemical space (see Methods and **SI Section 4**). This creates a molecular fingerprint of how elemental relationships are organized, showing how the use of heteroatoms (e.g., atoms that are not carbon or hydrogen, here including P, S, N, and O) varies with hydrogen content across Synthetic Chemical, Biochemical, and Environmental Microbial Spaces.

Across Synthetic Chemical, Biochemical, and Environmental Microbial Spaces, we observe systematic enrichment in heteroatoms with corresponding shifts in H:C, shown in **Figure 3**. Compounds in Synthetic Chemical Space contain relatively low proportions of heteroatoms relative to carbon across the chemical spaces. Synthetic Chemical Space, Biochemical Space, and Environmental Microbial Space all exhibited the largest spread in N:C and O:C ratios, while P:C and S:C ratios did not extend as far into the coordinate space. **Table 1** reports the average X (Element:C) and Y (H:C) coordinates of the highest-density regions, along with the $10^{th}$ and $90^{th}$ percentiles range of the data.

Phosphorus-containing compounds are relatively scarce in Synthetic Chemical Space compared with biochemical and Environmental Microbial Space. In Synthetic Chemical Space, the P:C values ranged from 0.03 – 0.17 with an average value of 0.09 and exhibited the highest corresponding H:C average of 1.42 for Synthetic Chemical Space with a range of 0.83 – 2.17. In biochemical and Environmental Microbial Space phosphorus-containing compounds are more common and more enriched in these elements compared to Synthetic Chemical Space. In Biochemical Space, the P:C values range from 0.04 – 0.25 with an average of 0.14, along with a relatively high associated H:C values ranging from 1.36 – 2.33 with an average of 1.78. In Environmental Microbial Space, the P:C values range from 0.04 – 0.30 with an average of 0.15, and associated H:C values range from 1.40 – 2.33 with an average of 1.79.

Sulfur exhibited similar scarcity and clustering with phosphorus in that it is rare to find compounds with high enrichments of P and S. In Synthetic Chemical Space the S:C ratios ranged from 0.03 – 0.13, with an average of 0.08. The corresponding H:C ratios exhibited an average of 1.12 and a range of 0.74 – 1.53. Biochemical space and microbial share similar and higher S:C enrichment compared to Synthetic Chemical Space. In Biochemical Space, S:C ratios range from 0.03 – 0.25 with an average of 0.13, with corresponding H:C values ranging from 0.89 – 2.00 and an average of 1.48. In Environmental Microbial Space, the S:C ratios range from 0.03 – 0.33 with an average of 0.13, and their corresponding H:C values range from 1.25 – 2.00 with an average of 1.66.

For Synthetic Chemical Space, the N:C values range from 0.05 – 0.29 with an average of 0.16. The associated H:C values range from 0.76 – 1.56 with an average of 1.14. Molecules in Biochemical Space contained higher proportions of heteroatoms relative to carbon. The N:C values range from 0.05 – 0.40 with an average of 0.20. The associated H:C values range from 0.89 – 2.00 with an average of 1.42. In Environmental Microbial Space, the N:C ratios exhibit the highest range from 0.07 – 0.50 with an average of 0.25. The associated H:C values range from 1.00 – 2.17 with an average of 1.59.

The O:C ratios in Synthetic Chemical Space are similar to those of N:C, with a range of 0.06 – 0.35 and an average of 0.19, with H:C values clustered near 1.16 with a range of 0.78 – 1.60. The O:C ratios in Biochemical Space are substantially higher than in Synthetic Chemical Space, ranging from 0.10 – 0.82 with an average of 0.40, and the corresponding H:C values range from 0.83 – 1.92 with an average of 1.38. Compounds in Environmental Microbial Space contain the highest proportions of heteroatoms and show elevated O:C ratios compared to the other molecular spaces. The O:C ratios in Environmental Microbial Space show the strongest enrichment, ranging from 0.13 – 1.19 with an average of 0.64, and the corresponding H:C values range from 1.00 – 2.17 with an average of 1.58.

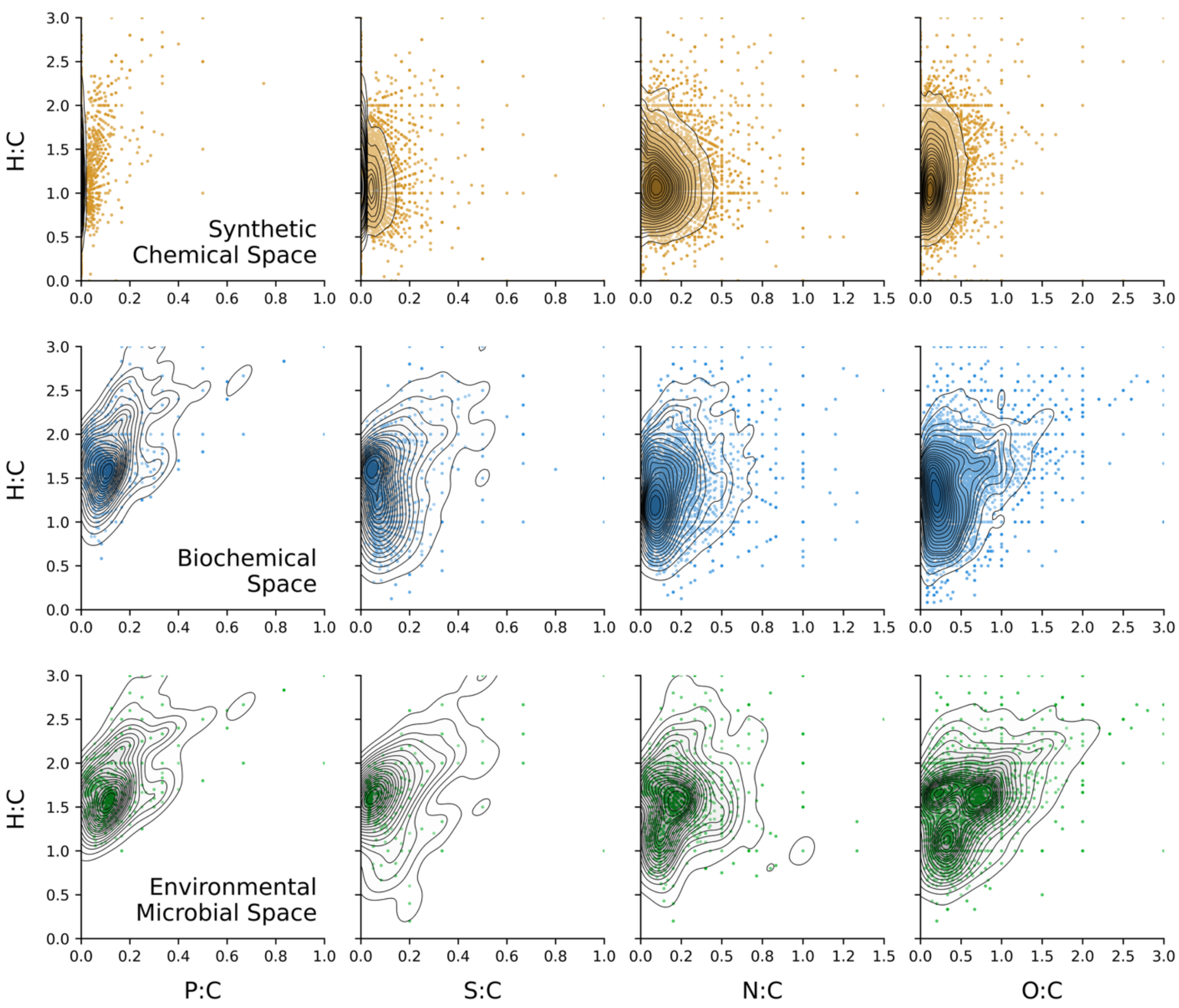

**Figure 3: Increasing heteroatom:C enrichment and clustering across P, S, N, and O chemical spaces.** Van Krevelen diagrams showing H:C ratio on the y-axis versus P:C, S:C, N:C, O:C ratios on the x-axis for three levels of chemical organization. Note that the x-axis scale differs between ratios. Each point represents the elemental composition of an individual compound, with grayscale contours indicating kernel density estimates. Rows correspond to (top) Synthetic Chemical Space, (middle) Biochemical Space, and (bottom) Environmental Microbial Space.

**Table 1: Summary statistics of Van Krevelen diagrams between Synthetic Chemical, Biochemical, and Environmental Microbial Spaces.** Average X (element:C) and Y (H:C) coordinates of kernel density estimations of compound clusters across chemical, biochemical, and Environmental Microbial Spaces, with $10^{th} – 90^{th}$ percentile ranges.

| | Ratio | Count | Avg. X:C | $10^{th}$ - $90^{th}$ % X:C | Avg. H:C | $10^{th}$ - $90^{th}$ % H:C |
|---|---|---|---|---|---|---|
| Synthetic Chemical Space | P:C | 584 | 0.09 | (0.03 - 0.17) | 1.42 | (0.83 - 2.17) |
| Biochemical Space | P:C | 2028 | 0.14 | (0.04 - 0.25) | 1.78 | (1.36 - 2.33) |
| Environmental Microbial Space | P:C | 904 | 0.15 | (0.04 - 0.30) | 1.79 | (1.40 - 2.33) |
| Synthetic Chemical Space | S:C | 4501 | 0.08 | (0.03 - 0.13) | 1.12 | (0.74 - 1.53) |
| Biochemical Space | S:C | 2321 | 0.13 | (0.03 - 0.25) | 1.48 | (0.89 - 2.00) |
| Environmental Microbial Space | S:C | 424 | 0.13 | (0.03 - 0.33) | 1.66 | (1.25 - 2.00) |
| Synthetic Chemical Space | N:C | 14365 | 0.16 | (0.05 - 0.29) | 1.14 | (0.76 - 1.56) |
| Biochemical Space | N:C | 9341 | 0.20 | (0.05 - 0.40) | 1.42 | (0.89 - 2.00) |
| Environmental Microbial Space | N:C | 1776 | 0.25 | (0.07 - 0.50) | 1.59 | (1.00 - 2.17) |
| Synthetic Chemical Space | O:C | 16265 | 0.19 | (0.06 - 0.35) | 1.16 | (0.78 - 1.60) |
| Biochemical Space | O:C | 18784 | 0.40 | (0.10 - 0.82) | 1.38 | (0.83 - 1.92) |
| Environmental Microbial Space | O:C | 3139 | 0.64 | (0.13 - 1.19) | 1.58 | (1.00 - 2.17) |

**Scaling of element ratios across systems**

To extend our element analysis from individual compounds across Synthetic Chemical, Biological, and Environmental Microbial Spaces to the element behavior of entire systems, we examined how element ratios change as a function of chemical space size. In this systems-level view, we compare the effect of Environmental Microbial Space, or enzymatic reactions inferred within community-level organization, with Synthetic Chemical Space compound sets. For each sample, we quantified the total counts of P, S, N, and O as a function of total C atoms in the metagenomic data (Environmental Microbial Space) and in the Synthetic Chemical Space compound sets, see **Figure 4**. We only compare scaling statistics between Environmental Microbial Space and Synthetic Chemical Space, allowing a comparison of element use in metabolic systems with the broader background of chemical space.

The element ratios within systems of molecules exhibit distinct scaling relationships between Environmental Microbial Space and Synthetic Chemical Space for P, S, N, and O relative to C. As the number of carbon atoms in the pool of molecules increases, the requirements for other elements scale with C according to distinct element ratio trajectories. Thus, scaling analyses capture emergent constraints on element use that arise only when many molecules are considered together, providing insight into systems-level organization that is not apparent from single-molecule stoichiometry alone.

Samples from Synthetic Chemical Space exhibit a linear scaling (slope = 1) relationship for N, O, and S. These elements scale proportionally with C, indicating the element-ratio requirements remain constant as the system size increases. Phosphorus in Synthetic Chemical Space exhibits superlinear scaling (slope > 1), indicating that the synthetic compound sets are becoming increasingly enriched with P as the amount of carbon within the compound sets increases. In Environmental Microbial Space, as communities increase in C atoms, the abundances of P, S, N, and O scale sublinearly with C (slope < 1), indicating a declining per-carbon demand for these elements. Phosphorus in Environmental Microbial Space exhibits the lowest sublinear scaling slope, revealing opposing behavior compared to Synthetic Chemical Space.

Environmental Microbial Space overall is more enriched in N, O, and P than Synthetic Chemical Space, with P displaying the most significant disparity in atom counts spanning 2 orders of magnitude between the groups. Unlike in N, O, and P, Synthetic Chemical Space and Environmental Microbial Space share similar values of S across system size. Synthetic Chemical Space overall can sustain more C and thus, more P, S, N, and O than Environmental Microbial Space.

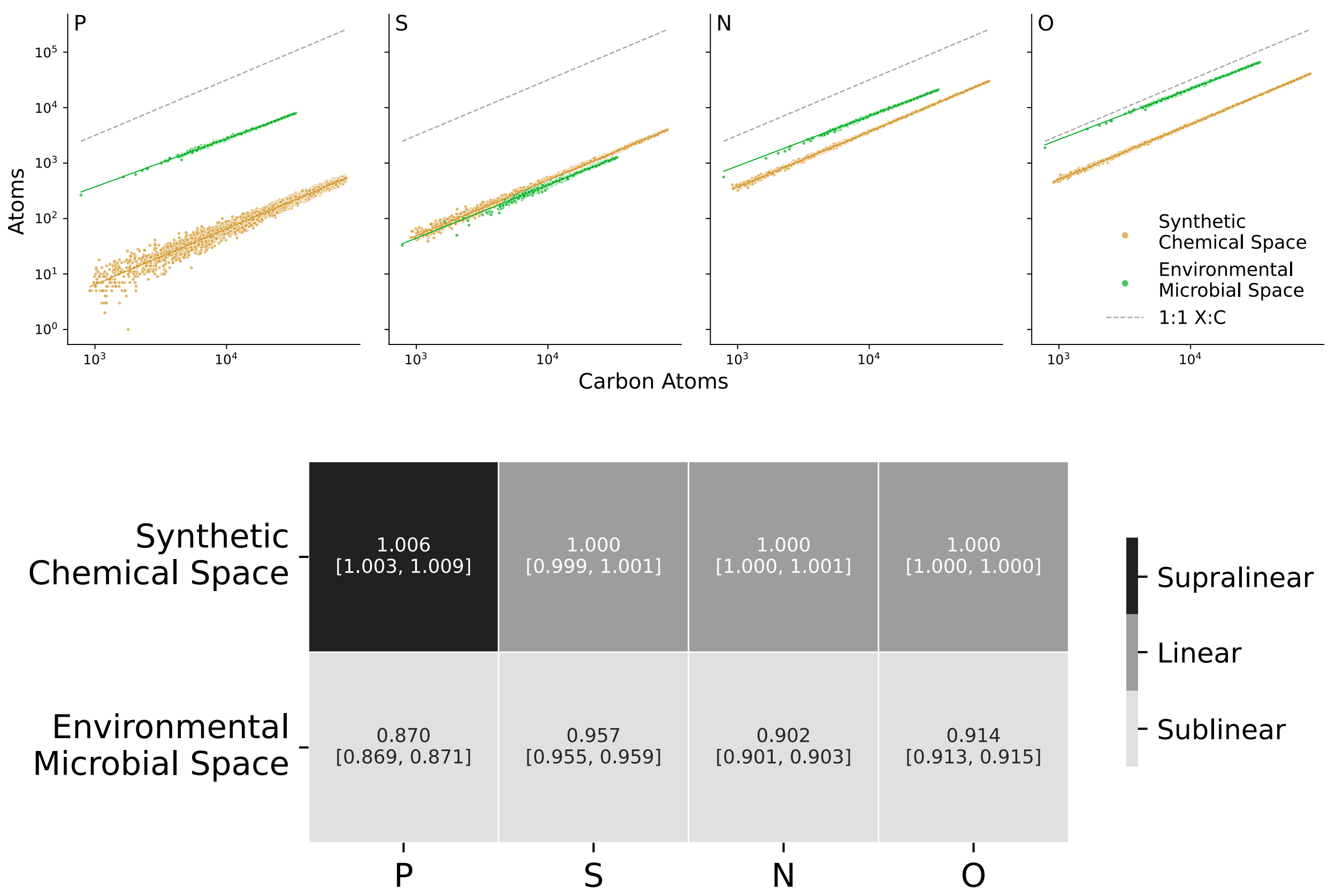

**Figure 4: Element scaling differentiates molecular data from environmentally sampled communities and compound sets sampled from Synthetic Chemical Space.** (A) Element scaling plots showing the total number of Phosphorus (P), Sulfur (S), Nitrogen (N), Oxygen (O) atoms across the molecules within each sample (y-axis) vs. the total number of Carbon (C) atoms across the molecules within each sample (x-axis). Each point represents the size of a community, either simulated (orange, Synthetic Chemical Space), or empirical (green, Environmental Microbial Space). (B) A heatmap showing the scaling slope values observed for P, S, N, and O as a function of C for Synthetic Chemical Space and Environmental Microbial Space with a 95% confidence interval.

### Towards ecological fingerprints in planetary data

We next explored the presence of element-ratio fingerprints in an extraterrestrial sample, the Bennu asteroid (OREX-803001-0), using the compounds detected through Gas Chromatography-Triple Quadrupole Mass Spectrometry (GC-QqQ-MS) and Liquid Chromatography-Time-of-Flight Mass Spectrometry (LC-FD/ToF-MS) from the hot water extracts. To contrast, we compared the element-ratio for P, S, N, and O in Van Krevelen diagrams between Environmental Microbial Space belonging to the ecosystem type “Rock-dwelling (endoliths),” hereafter “Endoliths,” since this could be considered an analog environment to an asteroid, see **Figure 5** and **Table 3**. We note that this comparison intended to demonstrate the methodology and is not a definitive test, which would require data standardization across both terrestrial and extraterrestrial samples, but we include it to motivate future work with more robust, standard data and larger sample sizes for developing statistical tests.

Endolithic metagenomes contain higher proportions of heteroatoms relative to carbon compared to Bennu compounds. Phosphorus-containing compounds are present in Endoliths with P:C values ranging from 0.04 – 0.30 and an average of 0.15, and associated H:C values ranging from 1.40 – 2.29 with an average of 1.76. In contrast, phosphorus-containing molecules were not detected in the defined extraction methods in the Bennu dataset, so the P:C value is set to 0. Endolithic S:C ratios range from 0.03 – 0.25 with an average of 0.11, with corresponding H:C values ranging from 1.31 – 2.00 and an average of 1.64. In Bennu, S:C ratios range from 0.21 – 0.32 with an average of 0.27, with corresponding H:C values ranging from 2.21 – 2.32 and an average of 2.27. The N:C values in Endoliths range from 0.07 – 0.50 with an average of 0.25, with associated H:C values ranging from 1.03 – 2.17 and an average of 1.58. In Bennu, N:C ratios span a similar range from 0.17 – 0.50 with an average of 0.32, but with higher associated H:C values ranging from 1.00 – 2.60 and an average of 1.97. The O:C ratios in Endoliths range from 0.14 – 1.20 with an average of 0.65, with corresponding H:C values ranging from 1.00 – 2.17 and an average of 1.59. In contrast, Bennu compounds exhibit lower O:C ratios ranging from 0.33 – 1.00 with an average of 0.54, with corresponding H:C values ranging from 1.00 – 2.33 and an average of 1.85.

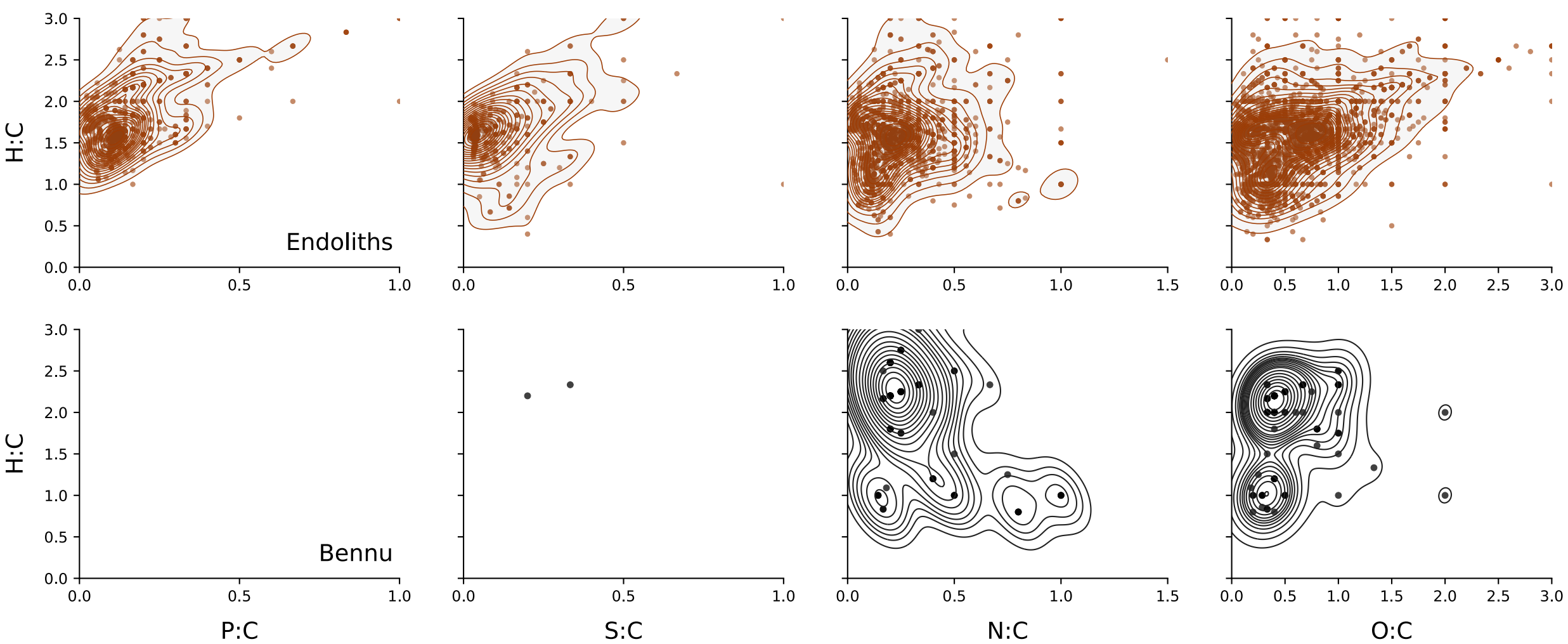

**Figure 5: Van Krevelen distributions of greater heteroatom enrichment and clustering in Endolith chemical space relative to Bennu.** Applications of Van Krevelen diagrams to metagenomes belonging to Endoliths ecosystem type (Top) and Bennu asteroid compound dataset (Bottom).

**Table 3: Summary statistics of Van Krevelen diagrams between Endolith and Bennu chemical spaces.** Average X (element:C) and Y (H:C) coordinates of kernel density estimations of compound clusters across the Environmental Microbial Space of metagenomes cataloged as Endoliths and the LC-MS detected compounds analyzed from Bennu sample OREX-803001-0, with 10th – 90th percentile ranges.

| | Ratio | Count | Avg. X:C | 10th - 90th % X:C | Avg. H:C | 10th - 90th % H:C |
|---|---|---|---|---|---|---|
| Endoliths | P:C | 803 | 0.15 | (0.04 - 0.30) | 1.76 | (1.40 - 2.29) |
| Bennu | P:C | 0 | 0.00 | (0.00 - 0.00) | 0.00 | (0.00 – 0.00) |
| Endoliths | S:C | 359 | 0.11 | (0.03 - 0.25) | 1.64 | (1.31 - 2.00) |
| Bennu | S:C | 2 | 0.27 | (0.21 - 0.32) | 2.27 | (2.21 - 2.32) |
| Endoliths | N:C | 1488 | 0.25 | (0.07 - 0.50) | 1.58 | (1.03 - 2.17) |
| Bennu | N:C | 83 | 0.32 | (0.17 - 0.50) | 1.97 | (1.00 - 2.60) |
| Endoliths | O:C | 2568 | 0.65 | (0.14 - 1.20) | 1.59 | (1.00 - 2.17) |
| Bennu | O:C | 91 | 0.54 | (0.33 - 1.00) | 1.85 | (1.00 - 2.33) |

# Discussion

### Environmental shaping of elemental use from Biochemical Space

We implemented a big data approach to demonstrate the utility of cheminformatic and element analyses to distinguish biologically selected sets of compounds: the long-term goal is applying this framework to diverse planetary samples to identify statistical fingerprints of ecological systems in molecular ensembles. Across our analyses, we observe a consistent increase in heteroatom:C ratios from Synthetic Chemical Space to Biochemical Space to Environmental Microbial Space. Environmental Microbial Space compounds are comparatively heteroatom-rich, or enriched in P, S, N, and O, compared to Synthetic Chemical and Biochemical Space (**Figure 2**). In Environmental Microbial Space, our data is a subset of compounds inferred to be used by the most abundant and metabolically diverse life on the planet (46). This increase in heteroatom:C ratios suggests that environmental microbial metabolism selects metabolites from bulk Biochemical Space that are enriched in P, S, N, and O with a particular functional organization.

The Redfield ratio is the atomic ratio of elements C:N:P found in biomass samples that generally follows the stoichiometric ratio of 106:16:1 (47). This ratio reflects phytoplankton biomass within an environmental sample. In chemical space, a C:N:P ratio does not reflect biomass abundance ratios in biological samples, rather it reflects the stoichiometric organization within chemical space. We calculate chemical space ratios based on the median ratio values in **Figure 2**, which correspond to C:N:P ratios of 17:2.2:1 in Synthetic Chemical Space, 9:1.3:1 in Biochemical Space, and 8.3:1.7:1 in Environmental Microbial Space. These ratios reflect not just biochemical structures, but the C:N:P of the small molecule chemistry selected by life. Although Redfield ratios and chemical space ratios are not directly comparable, the approaches provide corroborating evidence that living systems use more compounds enriched in C, H, P, S, N, and O relative to chemical space more broadly (47, 48, 25). A chemical space C:N:P ratio standard may be helpful in life detection when molecular abundance profiles are difficult to measure.

The progressive enrichment in heteroatom:C ratio from Synthetic Chemical to Biochemical and Environmental Microbial Space may be associated with a corresponding increase in clustering within Van Krevelen coordinate space. This is indicated by the kernel density estimation contours mapped in the coordinate space and summary statistics (**Figure 3**, **Table 1**). This increased enrichment pattern from Synthetic Chemical to Biochemical and to Environmental Microbial may reflect a fundamental difference in organization: Synthetic Chemical Space represents the general chemical structure of compounds without the influence of metabolic organization, Biochemical Space represents the compounds that support life broadly, whereas Environmental Microbial Space represents the subset of the compounds selected from Biochemical Space to use in enzyme catalyzed reactions shaped by functional and environmental constraints. Given how many of the compounds in Synthetic Chemical Space are synthetically derived by human chemists, they represent very different constraints to what Environmental Microbial life experiences, and therefore this contrasting data provides a useful reference dataset for determining methods for distinguishing data sources of origin based on elemental fingerprinting.

Taken together, the stoichiometry distribution and Van Krevelen fingerprints show differences between relatively less oxidized (H-saturated), heteroatom-poor molecules such as P, and S in Synthetic Chemical Space and more oxidized (H-unsaturated), heteroatom-enriched molecules in Biochemical Space and Environmental Microbial Space particularly in the case of oxygen, likely reflecting chemical and biochemical selection and environmental constraints observed within large-scale biomolecular data. This pattern may be due to the prevalence of these elements in specialized processes such as redox chemistry, energetic transfer mechanisms, or generating biomass (49–51). The more oxidized and heteroatom-enriched Van Krevelen diagrams may reflect how well life transforms resources into more oxidized molecules via metabolism.

**Ecological insights from element ratios across community scale systems**

We define the sublinear scaling observed in Environmental Microbial Space comprised of thousands of metagenomic communities as an ecosystem-scale feature of life's element use. The compound sets from Synthetic Chemical Space exhibit mostly linear relationships as system size increases while Environmental Microbial Space communities exhibit consistent sublinear scaling for the bioessential elements, P, S, N, and O. The sublinear scaling behavior shown by Environmental Microbial Space communities indicates a decrease in demand for compounds containing P, S, N, and O as the system size, or the number of C atoms increases. This suggests that larger communities are minimizing their P, S, N, and O use per C increase and functionality is maintained, indicating an increase in efficiency by the ability for the communities to synthesize or process novel compounds with more carbon and fewer of the other bioessential elements available for biochemistry.

Phosphorus shows the greatest deviation between Environmental Microbial Space Synthetic Chemical Space in terms of overall count values and the largest divergence in scaling behavior. This pattern is consistent with ecological stoichiometry principles: P is both essential and a limiting nutrient, constraining primary production across ecosystems. Phosphorus is central to energy transfer compounds such as ATP and is a central component of nucleic acids and cellular membranes (52). Unlike N, P cannot be fixed from an atmospheric reservoir, and its availability is ultimately governed by the chemical weathering of rocks and particular organic matter uptake, making it a long-term limiting nutrient in many environments (53). The strong P enrichment we see in Environmental Microbial Space likely reflects its indispensable role in metabolism despite its scarcity. The scaling divergence between microbial and Synthetic Chemical Space is consistent with how biological systems actively maintain P-rich molecules necessary to their function, whereas Synthetic Chemical Space lacks such constraints. As a result, the relative enrichment of P in a chemical system may serve as a sensitive indicator of biologically structured chemistry. In extraterrestrial environments, such as Martian soil or asteroid material, elevated phosphorus atoms relative to carbon could therefore provide evidence of metabolic processes shaping molecular composition (54). It is an open question whether P will be essential in alternative chemistries for life that do not depend on DNA or RNA, but the statistical pattern we observe across metabolites are not necessarily tied to the presence of DNA or RNA and allow the scope for using P-enrichment as a statistical biosignature in cases of biochemistry not like that on Earth. However, because P abundances are poorly constrained in both stellar and planetary systems, and remain difficult to measure observationally, our ability to assess this role remains limited (55).

Unlike P, S does not exhibit strong divergence between Microbial and Synthetic Chemical Space systems showing the comparatively most narrow range of the scaling fits. This feature might reflect the non-limiting nature of S in the environment, reducing the selective pressure for preferential incorporation into metabolism. Nevertheless, S plays essential roles in biochemistry, with origins tracing back to early metabolism, including its incorporation into amino acids such as cysteine and methionine and redox-active cofactors like iron-sulfur clusters (56). This dependence appears to have had a lasting effect on life's metabolism over evolutionary time, including a dependence on S in core metabolic processes such as its inclusion in Coenzyme A in the tricarboxylic acid cycle (TCA) cycle, in enzymatic reactions like transferases and nitrogenases, and in broad reaction processes such as methylation (57, 58). The absence of strong enrichment or scaling divergence despite these roles suggests that functional importance alone does not determine elemental representation in Environmental Microbial Space.

Nitrogen occupies a unique position in Environmental Microbial Space because its availability is largely mediated by life's ability to convert inert $N_2$ to bioavailable forms such as $NH_4^+$ with the nitrogenase enzyme (59). The functionality of nitrogenase is also highly dependent on the availability of metal co-factors such as Fe and Mo, linking N-fixation directly with environmental geochemistry (60, 61). Thus, N is a prime example of the strong coupling between metabolic function and environmental availability of resources. Although Environmental Microbial Space exhibits slightly higher overall count values, Synthetic Chemical Space N values are similar, reflecting the prevalence of N across compound diversity regardless of chemical origin. The difference in scaling behavior may be due to the unique functionality of N in metabolism. The sublinear scaling behavior observed here may reflect N-limitation in communities, where fixed N is redistributed through microbial interactions, including release by diazotrophs and rapid recycling of bioavailable N pools, reducing the need for de novo N-fixation as system size increases (62). Alternatively, the sublinear scaling may emerge from increased metabolic efficiency reflecting community-level partitioning of gene pathways in the nitrogen cycle (63).

Environmental Microbial Space O displays the highest count values in the scaling plots, in addition to reaching the highest heteroatom:C ratios. This may reflect the prevalence of O as a terminal electron acceptor in microbial metabolism. The rise of atmospheric oxygen during the Great

Oxidation Event fundamentally transformed Earth's chemical landscape, enabling the evolution of aerobic metabolism and dramatically increasing the energetic yield of biochemical reactions (64). This increase in oxidant availability expanded the diversity of energetically favorable metabolic pathways, enabling life to access a wider range of oxidized compounds and redox states (65). Despite the high O:C ratios observed, O also displays sublinear scaling with C in Environmental Microbial systems, suggesting that metabolic networks leverage the energetic advantages of O-based redox chemistry to support increasing chemical complexity without proportional increases in O demand per unit C. This pattern is consistent with oxygen's role in high-energy redox reactions, where large energetic returns from aerobic metabolism reduce the need for proportional increases in O incorporation as system size increases. Together, N and O highlight two distinct constraints on Environmental Microbial Space: N reflects limits on molecular composition, while O reflects roles in metabolism that are not directly tied to its incorporation into molecules.

The element scaling relationships suggest constraints in how elements are organized in Environmental Microbial Space. For example, one explanation for sublinear scaling behavior could be due to community resource sharing through symbiotic interactions such as mutualism (66, 67). Another possible explanation could be due to a conservation of mechanisms by selecting for optimized biochemical pathways through minimal steps in metabolism (68). It is possible that the sublinear behavior across Environmental Microbial Space is constrained by another variable not tested. Overall, elemental limitation appears to be a key factor shaping biochemical organization, with scarce elements such as P showing strong enrichment, while more available elements such as sulfur do not. This suggests that the presence of limitation, or scarcity of an element in a chemical system, may be helpful in defining universal features of evolving systems capable of detection. Our results suggest that sublinear scaling of bioessential elements, together with a Van Krevelen fingerprint characterized by heteroatom enrichment and increased clustering relative to Synthetic Chemical Space, may constitute a biosignature pair accessible through mass spectrometry, although more work will need to be done to determine if this pattern is truly universal, detectable directly from environmental samples, and if it could be applied in an extraterrestrial environment to identify life as we don't know it (69).

**Leveraging universal large-system scale element constraints for life detection**

As a proof of principle of the utility of this approach to such diverse datasets we considered whether element-ratio analyses of a subset of Environmental Microbial Space, Endolithic metagenomes, could be distinguished between ecosystems at large scales. Compounds inferred to be used by Endolithic life exhibit similar heteroatom-enriched clustering observed across Environmental Microbial Space with an exception in N:C ratios that are higher. We compare the compounds detected in the Bennu dataset (32, 33). Van Krevelen diagrams of the Ryugu asteroid, and Murchison, Tarda, and Orgueil meteorites show similar distributions (See **SI Section 7**).

The Bennu dataset occupies a broader and comparatively unstructured region of chemical space with higher H:C ratios for each element ratio and a reduced representation of P- and S-containing compounds. However, these samples are not a perfect, direct comparison due to differences in compound dataset acquisition. The observable chemical space of the Bennu dataset is constrained by the analytical methodology (33), just as the Endoliths are constrained by functional enzyme annotation and genetic sequencing. The molecules identified in asteroid and meteorite samples depend not only on what is present in the sample, but also on how the sample is analyzed. Different extraction methods, processing steps, and instrument sensitivities can change which compounds are detected and measured. Some methods recover more compounds, while others may miss or alter fragile molecules. Detection limits also determine which compounds can be confidently identified. As a result, the measured molecular diversity represents only the portion of compound diversity that is observable using a given analytical approach, rather than the complete chemical inventory of the sample.

Examining biochemistry through the lens of element organization in chemical space enables the identification of organizing principles based on physical and chemical constraints that may extend to chemistries of life elsewhere, or life as we don't know it. Although life on Earth uses enzymatic reactions for most of its chemical processes, life on another planet may operate under different functional constraints. Reaction diversity on Earth derived from the catalytic function of known enzymes, only relies on the chemical transformations involved in metabolism, and it is possible that patterns in those transformations could be generalizable to reactions mediated by other catalysts in an alternative chemistry for life. The patterns reported here capture statistical constraints on composition rather than specific compounds and may extend to metabolisms with different catalysts or functional requirements. Thus, investigating metabolism through the lens of chemical space allows for the potential to identify principles of biochemical organization. Large databases of meta-omics data offer an opportunity to build models that connect organismal processes with planetary-scale processes, with ecosystems acting as the bridge between the two, consistent with the view of ecosystems as complex adaptive systems in which large-scale patterns emerge from local interactions (70). We argue that this analysis provides generalizable insights into the emergent element organization patterns of metabolism as influenced by the environment. If such organizing principles are conserved, our framework investigating element stoichiometry may aid in predicting metabolic organization in environments beyond Earth.

Attempts to generalize whole organism stoichiometry beyond Earth life have revealed quite a large range in feasible elemental ratios for life (71). It is possible that generalizing the chemical space analysis done in this paper to alternative evolutionary trajectories could lead to different regions in **Figure 1**, and such regions might overlap differently with synthetic chemistry and Bennu samples. First-principles theories that would allow us to build such alternative spaces represent an exciting frontier in astrobiology.

# Conclusion

The results presented here demonstrate that elemental organization in chemical space exhibits statistical regularities, which can potentially differentiate between the source of origin, including whether a system is biologically derived or not. Together, sublinear elemental scaling and structured chemical space organization provide a framework for identifying ecological biosignatures based on small molecule elemental stoichiometry. Van Krevelen and elemental scaling analyses characterize both the structure of occupied chemical space and system-scale element use, requiring only molecular formula data and enabling direct application to mass spectrometric measurements from environmental samples, metabolomes, or planetary missions. This approach may offer a pathway toward identifying measurable ecological imprints in data from small molecule chemistry.

Acknowledgements

This work was supported by NASA Grant number 80NSSC21K1402. We thank the Emergence Walker laboratory for their aid, advice, and expertise in computational modeling and theory development.

## Author Contributions

Pilar Vergeli: conceptualization, methodology, software, validation, formal analysis, investigation, data curation, writing-original draft, writing – review & editing, visualization. Cole Mathis: conceptualization, data curation, supervision. John Malloy: data curation. L. Felipe Benites: data curation, writing – review & editing. Christopher P. Kempes: writing – review & editing. Elizabeth Trembath-Reichert: conceptualization, writing – review & editing, supervision. Hilairy Hartnett: conceptualization, methodology, writing – review & editing, supervision. Sara Walker: conceptualization, resources, writing – review & editing, supervision, project administration, funding acquisition.

# References

1. C. M. Dobson, Chemical space and biology. *Nature* **432**, 824–828 (2004).
2. W. Bains, S. Seager, A Combinatorial Approach to Biochemical Space: Description and Application to the Redox Distribution of Metabolism. *Astrobiology* **12**, 271–281 (2012).
3. J.-L. Reymond, The Chemical Space Project. *Acc. Chem. Res.* **48**, 722–730 (2015).
4. R. S. Bohacek, C. McMartin, W. C. Guida, The art and practice of structure-based drug design: A molecular modeling perspective. *Med. Res. Rev.* **16**, 3–50 (1996).
5. M. Kanehisa, S. Goto, KEGG: Kyoto Encyclopedia of Genes and Genomes. *Nucleic Acids Res.* **28**, 27–30 (2000).
6. S. M. Marshall, *et al.*, Identifying molecules as biosignatures with assembly theory and mass spectrometry. *Nat. Commun.* **12**, 3033 (2021).
7. M. Jirasek, *et al.*, Investigating and Quantifying Molecular Complexity Using Assembly Theory and Spectroscopy. *ACS Cent. Sci.* **10**, 1054–1064 (2024).
8. L. Cronin, S. I. Walker, The Physics of Causation. [Preprint] (2026). Available at: http://arxiv.org/abs/2601.00515 [Accessed 13 May 2026].
9. L. Seyler, *et al.*, Metabolomics as an Emerging Tool in the Search for Astrobiologically Relevant Biomarkers. *Astrobiology* **20**, 1251–1261 (2020).
10. L. Chou, *et al.*, Planetary Mass Spectrometry for Agnostic Life Detection in the Solar System. *Front. Astron. Space Sci.* **8** (2021).

11. G. M. Weiss, *et al.*, Operational considerations for approximating molecular assembly by Fourier transform mass spectrometry. *Front. Astron. Space Sci.* **11** (2024).

12. L. A. Rutter, *et al.*, Exploring molecular assembly as a biosignature using mass spectrometry and machine learning. [Preprint] (2025). Available at: http://arxiv.org/abs/2507.19057 [Accessed 23 April 2026].

13. H. J. Cleaves, *et al.*, A robust, agnostic molecular biosignature based on machine learning. *Proc. Natl. Acad. Sci.* **120**, e2307149120 (2023).

14. M. L. Wong, *et al.*, Organic geochemical evidence for life in Archean rocks identified by pyrolysis-GC-MS and supervised machine learning. *Proc. Natl. Acad. Sci. U. S. A.* **122**, e2514534122 (2025).

15. G. Yoffe, F. Klenner, B. Sober, Y. Kaspi, I. Halevy, Molecular diversity as a biosignature. *Nat. Astron.* 1–10 (2026). https://doi.org/10.1038/s41550-026-02864-z.

16. J. J. R. Fraústro da Silva, R. J. P. Williams, *The Biological Chemistry of the Elements: The Inorganic Chemistry of Life* (OUP Oxford, 2001).

17. J. J. Elser, C. Acquisti, S. Kumar, Stoichiogenomics: the evolutionary ecology of macromolecular elemental composition. *Trends Ecol. Evol.* **26**, 38–44 (2011).

18. E. L. Shock, E. S. Boyd, Principles of Geobiochemistry. *Elements* **11**, 395–401 (2015).

19. K. A. Remick, J. D. Helmann, "Chapter One - The elements of life: A biocentric tour of the periodic table" in *Advances in Microbial Physiology*, R. K. Poole, D. J. Kelly, Eds. (Academic Press, 2023), pp. 1–127.

20. N. R. Pace, The universal nature of biochemistry. *Proc. Natl. Acad. Sci.* **98**, 805–808 (2001).

21. S. L. Schreiber, The small-molecule approach to biology. *Chem Eng News* **81**, 51–61 (2003).

22. S. Seager, W. Bains, J. J. Petkowski, Toward a List of Molecules as Potential Biosignature Gases for the Search for Life on Exoplanets and Applications to Terrestrial Biochemistry. *Astrobiology* **16**, 465–485 (2016).

23. S. I. Walker, *et al.*, Exoplanet Biosignatures: Future Directions. *Astrobiology* **18**, 779–824 (2018).

24. J. Sardans, A. Rivas-Ubach, J. Peñuelas, The elemental stoichiometry of aquatic and terrestrial ecosystems and its relationships with organismic lifestyle and ecosystem structure and function: a review and perspectives. *Biogeochemistry* **111**, 1–39 (2012).

25. J. Peñuelas, *et al.*, The bioelements, the elementome, and the biogeochemical niche. *Ecology* **100**, e02652 (2019).

26. D. Van Krevelen, Graphical-statistical method for the study of structure and reaction processes of coal. *Fuel* **29**, 269–284 (1950).

27. A. Rivas-Ubach, *et al.*, Moving beyond the van Krevelen Diagram: A New Stoichiometric Approach for Compound Classification in Organisms. *Anal. Chem.* **90**, 6152–6160 (2018).

28. C. P. Kempes, *et al.*, Generalized Stoichiometry and Biogeochemistry for Astrobiological Applications. *Bull. Math. Biol.* **83**, 73 (2021).

29. A. J. Lawson, J. Swienty-Busch, T. Géoui, D. Evans, "The Making of Reaxys—Towards Unobstructed Access to Relevant Chemistry Information" in *ACS Symposium Series*, L. R. McEwen, R. E. Buntrock, Eds. (American Chemical Society, 2014), pp. 127–148.

30. A. Clum, *et al.*, DOE JGI Metagenome Workflow. *mSystems* **6**, e00804-20 (2021).

31. L. Slocombe, C. Millsaps, K. Narasimhan, S. Walker, CBRdb: A Curated Biochemical Reaction Database for Precise Biochemical Reaction Analysis. [Preprint] (2025). Available at: https://chemrxiv.org/engage/chemrxiv/article-details/67c28c046dde43c908f7aa37 [Accessed 3 June 2025].

32. D. P. Glavin, *et al.*, Abundant ammonia and nitrogen-rich soluble organic matter in samples from asteroid (101955) Bennu. *Nat. Astron.* **9**, 199–210 (2025).

33. Y. Oba, *et al.*, Distribution of extraterrestrial nucleobases, other N-heterocycles, and their precursors in a sample from asteroid Bennu. *Commun. Chem.* **9**, 132 (2026).

34. M. Huntemann, *et al.*, The standard operating procedure of the DOE-JGI Metagenome Annotation Pipeline (MAP v.4). *Stand. Genomic Sci.* **11**, 17 (2016).

35. I.-M. A. Chen, *et al.*, IMG/M v.5.0: an integrated data management and comparative analysis system for microbial genomes and microbiomes. *Nucleic Acids Res.* **47**, D666–D677 (2019).

36. D. C. Gagler, *et al.*, Scaling laws in enzyme function reveal a new kind of biochemical universality. *Proc. Natl. Acad. Sci.* **119**, e2106655119 (2022).

37. A. G. McDonald, K. F. Tipton, Enzyme nomenclature and classification: the state of the art. *FEBS J.* **290**, 2214–2231 (2023).

38. S. J. Giovannoni, *et al.*, Genome Streamlining in a Cosmopolitan Oceanic Bacterium. *Science* **309**, 1242–1245 (2005).

39. S. Mukherjee, *et al.*, Genomes OnLine Database (GOLD) v.10: new features and updates. *Nucleic Acids Res.* **53**, D989–D997 (2025).

40. D. I. Osolodkin, *et al.*, Progress in visual representations of chemical space. *Expert Opin. Drug Discov.* **10**, 959–973 (2015).

41. C. L. Heald, *et al.*, A simplified description of the evolution of organic aerosol composition in the atmosphere. *Geophys. Res. Lett.* **37** (2010).

42. H.-Y. Li, *et al.*, The chemodiversity of paddy soil dissolved organic matter correlates with microbial community at continental scales. *Microbiome* **6**, 187 (2018).

43. P. Virtanen, *et al.*, SciPy 1.0: fundamental algorithms for scientific computing in Python. *Nat. Methods* **17**, 261–272 (2020).

44. T. Fink, H. Bruggesser, J.-L. Reymond, Virtual exploration of the small-molecule chemical universe below 160 daltons. *Angew. Chem. Int. Ed.* **44**, 1504–1508 (2005).

45. J.-L. Reymond, L. Ruddigkeit, L. Blum, R. van Deursen, The enumeration of chemical space. *WIREs Comput. Mol. Sci.* **2**, 717–733 (2012).

46. P. G. Falkowski, T. Fenchel, E. F. Delong, The Microbial Engines That Drive Earth's Biogeochemical Cycles. *Science* **320**, 1034–1039 (2008).

47. A. C. Redfield, The Biological Controls of Chemical Factors in the Environment. *Am. Sci.* **46**, 230A, 205–221 (1958).

48. J. J. Elser, Biological stoichiometry: a theoretical framework connecting ecosystem ecology, evolution, and biochemistry for application in astrobiology. *Int. J. Astrobiol.* **2**, 185–193 (2003).

49. R. W. Sterner, J. J. Elser, Ecological Stoichiomety: The biology of elements from molecules to the biosphere Princeton University Press. *Princet. N. J. USAGoogle Sch.* (2002).

50. A. J. Burgin, W. H. Yang, S. K. Hamilton, W. L. Silver, Beyond carbon and nitrogen: how the microbial energy economy couples elemental cycles in diverse ecosystems. *Front. Ecol. Environ.* **9**, 44–52 (2011).

51. S. Ramírez-Flandes, B. González, O. Ulloa, Redox traits characterize the organization of global microbial communities. *Proc. Natl. Acad. Sci.* **116**, 3630–3635 (2019).

52. A. Paytan, K. McLaughlin, The Oceanic Phosphorus Cycle. *Chem. Rev.* **107**, 563–576 (2007).

53. P. M. Vitousek, S. Porder, B. Z. Houlton, O. A. Chadwick, Terrestrial phosphorus limitation: mechanisms, implications, and nitrogen–phosphorus interactions. *Ecol. Appl.* **20**, 5–15 (2010).

54. M. Fernández-Ruz, I. Jiménez-Serra, J. Aguirre, A Theoretical Approach to the Complex Chemical Evolution of Phosphorus in the Interstellar Medium. *Astrophys. J.* **956**, 47 (2023).

55. N. R. Hinkel, H. E. Hartnett, P. A. Young, The Influence of Stellar Phosphorus on Our Understanding of Exoplanets and Astrobiology. *Astrophys. J.* **900**, L38 (2020).

56. M. C. Weiss, M. Preiner, J. C. Xavier, V. Zimorski, W. F. Martin, The last universal common ancestor between ancient Earth chemistry and the onset of genetics. *PLOS Genet.* **14**, e1007518 (2018).

57. J. E. Goldford, H. Hartman, R. Marsland, D. Segrè, Environmental boundary conditions for the origin of life converge to an organo-sulfur metabolism. *Nat. Ecol. Evol.* **3**, 1715–1724 (2019).

58. A. Francioso, A. Baseggio Conrado, L. Mosca, M. Fontana, Chemistry and Biochemistry of Sulfur Natural Compounds: Key Intermediates of Metabolism and Redox Biology. *Oxid. Med. Cell. Longev.* **2020**, 8294158 (2020).

59. J. Raymond, J. L. Siefert, C. R. Staples, R. E. Blankenship, The Natural History of Nitrogen Fixation. *Mol. Biol. Evol.* **21**, 541–554 (2004).

60. A. Anbar, A. Knoll, Proterozoic Ocean Chemistry and Evolution: A Bioinorganic Bridge? *Science* **297**, 1137–1142 (2002).

61. J. B. Glass, F. Wolfe-Simon, A. D. Anbar, Coevolution of metal availability and nitrogen assimilation in cyanobacteria and algae. *Geobiology* **7**, 100–123 (2009).

62. M. R. Mulholland, The fate of nitrogen fixed by diazotrophs in the ocean. *Biogeosciences* **4**, 37–51 (2007).

63. A. V. Carr, *et al.*, Emergence and disruption of cooperativity in a denitrifying microbial community. [Preprint] (2024). Available at: http://biorxiv.org/lookup/doi/10.1101/2024.10.24.620115 [Accessed 12 March 2026].

64. T. W. Lyons, C. T. Reinhard, N. J. Planavsky, The rise of oxygen in Earth's early ocean and atmosphere. *Nature* **506**, 307–315 (2014).

65. T. W. Lyons, *et al.*, Co-evolution of early Earth environments and microbial life. *Nat. Rev. Microbiol.* **22**, 572–586 (2024).

66. R. E. Turner, Element ratios and aquatic food webs. *Estuaries* **25**, 694–703 (2002).

67. S. Schmidt, *et al.*, From soil to symbiosis: elemental filtering in a termite-fungus mutualism. *Soil Biol. Biochem.* **213**, 110045 (2026).

68. E. Noor, E. Eden, R. Milo, U. Alon, Central Carbon Metabolism as a Minimal Biochemical Walk between Precursors for Biomass and Energy. *Mol. Cell* **39**, 809–820 (2010).

69. S. I. Walker, *Life as no one knows it: the physics of life's emergence* (Penguin, 2024).

70. S. A. Levin, Ecosystems and the Biosphere as Complex Adaptive Systems. *Ecosystems* **1**, 431–436 (1998).

71. D. Muratore, S. I. Walker, H. Graham, C. H. House, C. P. Kempes, “Observations of Elemental Composition of Enceladus Consistent with Generalized Models of Theoretical Ecosystems” (Microbiology, 2023).